\documentclass{IEEEtran}

\usepackage{cite}

\ifCLASSINFOpdf
\usepackage[pdftex]{graphicx}
\else
\usepackage[dvips]{graphicx}
\fi

\usepackage{caption}
\usepackage{subcaption}

\usepackage[cmex10]{amsmath}
\usepackage{amsfonts, amssymb, amsthm}
\usepackage{bm}
\usepackage{physics}

\newcommand{\pcec}[2]{\ensuremath{\mathsf{#1}^{#2}}}
\newcommand{\expv}[1]{\ensuremath{\mathbb{E}\left[ #1 \right]}}
\newcommand{\varv}[1]{\ensuremath{\mathbb{V}\left[ #1 \right]}}

\newtheorem{definition}{Definition}

\newtheorem{remark}{Remark}

\usepackage{algorithmic}
\usepackage{algorithm}

\usepackage{array}
\usepackage{multirow}

\usepackage[hidelinks]{hyperref}
\usepackage{orcidlink}

\usepackage{xcolor}

\begin{document}
    \title{Towards Stochastic (N-1)-Secure Redispatch
    \thanks{OM and TF are with the Institute of Control Systems, Hamburg University
    of Technology, Hamburg, Germany {\tt\small oleksii.molodchyk@tuhh.de} and {\tt\small timm.faulwasser@ieee.org}.
    HD is with the Department of Mathematics, TU Dortmund University, Dortmund, Germany
    {\tt\small hendrik.droegehorn@tu-dortmund.de}. ML and MK are with 50Hertz
    Transmission GmbH, Berlin, Germany
    {\tt\small \{martin.lindner, mario.kendziorski\}@50hertz.com}.
    \newline
    \noindent
    OM and TF acknowledge funding in the course of TRR 391 \textit{Spatio-temporal
    Statistics for the Transition of Energy and Transport} (520388526) by the
    Deutsche Forschungsgemeinschaft (DFG).}
    }

    \author{Oleksii Molodchyk\,\orcidlink{0009-0001-1659-1891}, Hendrik
    Drögehorn\,\orcidlink{0009-0004-5453-4750}, Martin Lindner\,\orcidlink{0000-0001-8466-2666},
    Mario Kendziorski\,\orcidlink{0000-0001-5508-2953}, Timm Faulwasser\,\orcidlink{0000-0002-6892-7406}}

    \maketitle

    \begin{abstract}
        The intermittent nature of renewable power availability is one of the major
        sources of uncertainty in power systems. While markets can guarantee
        that the demand is covered by the available generation, transmission system
        operators have to often intervene via economic redispatch to ensure that
        the physical constraints of the network are satisfied. To account for
        uncertainty, the underlying optimal power flow (OPF) routines have to be
        modified. Recently, polynomial chaos expansion (PCE) has been suggested in
        the literature as a tool for stochastic OPF problems. However, the usage
        of PCE-based methods in security-constrained OPF for (N-1)-secure operations
        has not yet been explored. In this paper, we propose a procedure that iteratively
        solves a PCE-overloaded stochastic OPF problem by including line outage constraints
        until an (N-1)-secure solution is achieved. We demonstrate the efficacy
        of our method by comparing it with a Monte-Carlo simulation on a 118-bus
        example system.
    \end{abstract}

    \section{Introduction}
    \label{sec:intro} In the European zonal market, congestion management is a
    key responsibility for Transmission System Operators (TSOs), who must ensure
    the physical feasibility of market outcomes. When necessary, TSOs implement redispatch
    measures---geographically reallocating scheduled generation---to maintain (N-1)-secure
    grid operation. In 2024, redispatch costs in Germany totaled \texteuro 2.9 billion
    \cite{BNetzA_CongestionManagement_2022}.

    Minimizing these costs can be formulated as a Security-Constrained Optimal Power
    Flow (SCOPF) problem \cite{capitanescuStateoftheartChallengesFuture2011},
    solved repeatedly by TSOs during day-ahead and intraday operational planning
    using the latest forecasts and schedules for generation and demand. Currently,
    these time series are treated as deterministic inputs, lacking information about
    the variability/uncertainty. Consequently, redispatch measures are also deterministic
    and require manual adjustment by operators in response to real-time load flow
    situations.

    SCOPF incorporates static security margins to address load flow uncertainties
    between planning and real-time operation. However, this approach is limited:
    it does not guarantee that redispatch measures will be sufficient to
    mitigate congestion in real-time, nor does it identify grid areas with high power
    flow volatility. In grids dominated by Renewable Energy Sources (RES), securing
    limited ramp-up potential is crucial, whereas for conventional power plants,
    one must consider lead times of up to several hours. This necessitates
    uncertainty-aware decision making well before real-time operation.

    As a result, there is a growing need for computationally efficient stochastic
    redispatch optimization frameworks that enable informed operational planning
    under uncertainty. These tools must also meet the hard timing requirements
    of daily operations. Following recent literature, we rely on random-variable
    representations using polynomial chaos expansion (PCE), originally introduced
    by Norbert Wiener~\cite{wienerHomogeneousChaos1938}, to include stochastic
    uncertainties into the optimization.

    One of the appealing properties of PCEs is the possibility to analyze the
    contribution of individual uncertainties to any output of the model
    \cite{sudretGlobalSensitivityAnalysis2008}. This has motivated their use in stochastic
    DC or AC power flow and optimal power flow (OPF) applications \cite{niBasisAdaptiveSparsePolynomial2017,muhlpfordtGeneralizedFrameworkChanceconstrained2018,
    muhlpfordtChanceConstrainedACOptimal2019}. While there are approaches to
    include (N-1)-security considerations into stochastic OPF~\cite{sharifzadehStochasticSecurityconstrainedOptimal2016,
    sundarChanceConstrainedUnitCommitment2019}, to the best of the authors'
    knowledge, the use of PCEs for stochastic security-constrained OPF with (N-1)-security
    has not yet been considered, even though PCEs can be applied for
    probabilistic risk assessment under line outages
    \cite{lyNovelQuantileLitePCE2023}. In this paper, we provide an algorithm
    for stochastic redispatch based on (N-1)-SCOPF, wherein the uncertain wind and
    solar forecasts are modeled using beta distributions that can be fitted to mean
    and quantile data.

    The remainder of this paper is structured as follows. Section~\ref{sec:basics}
    introduces a generic deterministic OPF problem and explains how PCE-based
    descriptions of random variables can be included into the optimization to
    account for the uncertainties. An algorithm to combine the resulting stochastic
    optimization problem with the (N-1)-security criterion is provided in Section~\ref{sec:main_results}.
    Subsequently, this algorithm is tested on a IEEE 118-bus system, where it is
    compared with the Monte-Carlo (MC) approach, where a deterministic (N-1)-security
    constrained OPF instance is solved for each forecast realization (scenario).
    Finally, Section~\ref{sec:conclusion} concludes the paper and provides some directions
    for future work.

    \subsubsection*{Notation}
    For some $n \in \mathbb{N}$, we denote the set $\{1, \ldots, n\}$ as
    $\mathbb{I}_{n}$ and we use the shorthand
    $\mathbb{N}_{0}\doteq \mathbb{N}\cup \{0\}$. For a set $\mathcal{A}$, we
    denote the number of its elements by $\abs{\mathcal{A}}$. For a vector $x \in
    \mathbb{R}^{\abs{\mathcal{A}}}$, whose entries are indexed by the elements
    of $\mathcal{A}$, for each $i \in \mathcal{A}$, the $i$-th entry of $x$ is
    denoted by $x_{i}$. Similarly, for a matrix
    $\mathbf{X}\in \mathbb{R}^{\abs{\mathcal{A}} \times \abs{\mathcal{A}}}$
    indexed by $\mathcal{A}\times \mathcal{A}$, this carries over as $\mathbf{X}_{ij}$
    for $(i,j) \in \mathcal{A}\times \mathcal{A}$.

    \section{Preliminaries}
    \label{sec:basics} We consider a balanced electric power network at steady state
    with the set of buses $\mathcal{N}$ connected by branches (transmission
    lines or transformers) $\mathcal{E}\subset \mathcal{N}\times \mathcal{N}$.
    Let each $ij \doteq (i ,j) \in \mathcal{E}$ be equipped with a positive
    reactance and assume that the power system contains a set of conventional
    generators $\mathcal{G}$ and a set of RES $\mathcal{R}$; let the total power
    supplied by $\mathcal{G}\cup \mathcal{R}$ be consumed by the demands indexed
    with the set $\mathcal{D}$.

    We consider the DC power flow model \cite{stottDCPowerFlow2009} which reads
    \begin{subequations}
        \label{eq:dc_pf}
        \begin{align}
            \label{eq:conservation} & \sum_{i \in \mathcal{G}}p_{\mathrm{G},i}+ \sum_{i \in \mathcal{R}}p_{\mathrm{R},i}- \sum_{i \in \mathcal{D}}p_{\mathrm{D},i}= 0,                                \\
            \label{eq:power_flows}  & p_{\mathrm{f}}= \mathrm{PTDF}\left( \mathbf{C}_{\mathrm{G}}p_{\mathrm{G}}+ \mathbf{C}_{\mathrm{R}}p_{\mathrm{R}}- \mathbf{C}_{\mathrm{D}}p_{\mathrm{D}}\right).
        \end{align}
    \end{subequations}
    Here, $p_{\mathrm{D}}\in \mathbb{R}^{\abs{\mathcal{D}}}$ is the vector of
    all active power demands, whereas $p_{\mathrm{G}}\in \mathbb{R}^{\abs{\mathcal{G}}}$
    and $p_{\mathrm{R}}\in \mathbb{R}^{\abs{\mathcal{R}}}$ denote all active power
    generations of the units in $\mathcal{G}$ and $\mathcal{R}$, respectively. We
    use $\mathbf{C}_{\mathrm{G}}\in \lbrace 0,1 \rbrace^{\abs{\mathcal{N}} \times
    \abs{\mathcal{G}}}$ to map the generators $\mathcal{G}$ to the buses by setting
    $\mathbf{C}_{\mathrm{G},ij} \doteq 1$ if $j \in \mathcal{G}$ is located at $i
    \in \mathcal{N}$ and $\mathbf{C}_{\mathrm{G},ij} \doteq 0$ otherwise. The
    matrices
    $\mathbf{C}_{\mathrm{R}}\in \lbrace 0,1 \rbrace^{\abs{\mathcal{N}} \times \abs{\mathcal{R}}}$
    and
    $\mathbf{C}_{\mathrm{D}}\in \lbrace 0,1 \rbrace^{\abs{\mathcal{N}} \times \abs{\mathcal{D}}}$
    are defined analogously to $\mathbf{C}_{\mathrm{G}}$ for the RES and demands,
    respectively. Finally, the $\abs{\mathcal{E}}\times \abs{\mathcal{N}}$ power
    transfer distribution factor (PTDF) matrix maps the net nodal active power injections
    to the active power flows $p_{\mathrm{f}}\in \mathbb{R}^{\abs{\mathcal{E}}}$
    on all branches. We assume that the TSO can control all units in
    $\mathcal{G}\cup \mathcal{R}$ and that the demands $\mathcal{D}$ are
    uncontrollable but known.

    \subsection{The Redispatch Problem and (N-1)-Security}
    We focus on a market clearing scenario in which the power demand
    $p_{\mathrm{D}}$ has been matched by the power supplies
    $p_{\mathrm{G}}^{\bullet}$ and $p_{\mathrm{R}}^{\bullet}$ from the
    conventional and renewable units, respectively and such that \eqref{eq:conservation}
    is satisfied. To maintain a feasible operating point, the TSO can adjust the
    base feed-ins $p_{\mathrm{G}}^{\bullet}, p_{\mathrm{R}}^{\bullet}$ by
    solving an active power redispatch problem
    \begin{subequations}
        \label{eq:redispatch}
        \begin{alignat}
            {4} & \underset{p_{\mathrm{G}}^+, p_{\mathrm{G}}^-, p_{\mathrm{R}}^-}{\text{minimize}}~~ &  & f(p_{\mathrm{G}}^{+}, p_{\mathrm{G}}^{-}, p_{\mathrm{R}}^{-}) \label{eq:red_objective}                                                                                   \\
                & ~~~~~~~\text{s.t.}~~                                                               &  & p_{\mathrm{G}}= p_{\mathrm{G}}^{\bullet}+ p_{\mathrm{G}}^{+}- p_{\mathrm{G}}^{-}, ~~p_{\mathrm{R}}= p_{\mathrm{R}}^{\bullet}- p_{\mathrm{R}}^{-}\label{eq:red_pg_update} \\
                &                                                                                    &  & \text{DC power flow \eqref{eq:dc_pf}}, \notag                                                                                                                            \\
                &                                                                                    &  & 0 \leq p_{\mathrm{G},i}^{s}\leq p_{\mathrm{G},i}^{s,\max}, \quad \forall \, s \in \{+,-\}, \forall \, i \in \mathcal{G}, \label{eq:red_pg_con_p}                         \\
                &                                                                                    &  & 0 \leq p_{\mathrm{R},j}^{-}\leq p_{\mathrm{R},j}^{-,\max}, \quad \forall \, j \in \mathcal{R}, \label{eq:red_pr_con}                                                     \\
                &                                                                                    &  & \abs{p_{\mathrm{f}, km}}\leq p_{\mathrm{f}, km}^{\max}, \quad \forall \, (k,m) \in \mathcal{E}.\label{eq:red_pf_con}
        \end{alignat}
    \end{subequations}
    Here, equations \eqref{eq:red_pg_update} represent the redispatch measures. Conventional
    units can be ramped up by $p_{\mathrm{G}}^{+}\in \mathbb{R}^{\abs{\mathcal{G}}}$
    or down by $p_{\mathrm{G}}^{-}\in \mathbb{R}^{\abs{\mathcal{G}}}$, whereas
    RES can only be curtailed by
    $p_{\mathrm{R}}^{-}\in \mathbb{R}^{\abs{\mathcal{R}}}$. All redispatch measures
    are box-constrained via \eqref{eq:red_pg_con_p}-\eqref{eq:red_pr_con}. In
    \eqref{eq:red_pf_con}, the power flow over each $(k,m) \in \mathcal{E}$ is
    limited by its maximum allowable value $p_{\mathrm{f}, km}^{\max}$ in both
    directions. Finally, we set up the objective \eqref{eq:red_objective} as a sum
    \begin{subequations}
        \label{eq:objective}
        \begin{equation}
            \label{eq:objective1}f(p_{\mathrm{G}}^{+}, p_{\mathrm{G}}^{-}, p_{\mathrm{R}}
            ^{-}) \doteq \textstyle \sum\limits_{i \in \mathcal{G}}g_{i}(p_{\mathrm{G},i}
            ^{+}, p_{\mathrm{G},i}^{-}) + \textstyle\sum\limits_{j \in
            \mathcal{R}}r_{j}(p_{\mathrm{R},j}^{-})
        \end{equation}
        of convex quadratic costs $g_{i}: \mathbb{R}^{2}\to \mathbb{R}$ and $r_{j}
        : \mathbb{R}\to \mathbb{R}$ with
        \begin{equation}
            \label{eq:objective2}
            \begin{aligned}
                g_{i}(p,p^{\prime}) & \doteq g_{2,i}^{+}p^{2}+ g_{1,i}^{+}p+g_{2,i}^{-}(p^{\prime})^{2}+ g_{1,i}^{-}(p^{\prime}), \\
                r_{j}(p)            & \doteq r_{2,j}p^{2}+ r_{1,j}p,
            \end{aligned}
        \end{equation}
    \end{subequations}
    and with scalar coefficients $g_{k,i}^{+}$, $g_{k,i}^{-}$, and $r_{k,j}$ for
    $k \in \{1,2\}$. Thus, \eqref{eq:redispatch} is a convex quadratic
    programming problem.

    To choose a more secure operating point, TSOs often consider the (N-1)-security
    criterion, outlined below.
    \begin{definition}[(N-1)-Security~\cite{endrenyiBulkPowerSystemReliability1988}]
        \label{def:n1_security} An operating point specified by $(p_{\mathrm{G}},
        p_{\mathrm{R}})$ is considered (N-1)-secure if the power flows \eqref{eq:power_flows}
        satisfy the constraints \eqref{eq:red_pf_con} not only in the intact system
        but also after the outage of any branch $(k ,m) \in \mathcal{E}$.
    \end{definition}

    The above definition can be generalized to include outages of other
    components such as generators, switches, etc~\cite{endrenyiBulkPowerSystemReliability1988}.
    In this paper, however, we focus on branch outages only.

    Consider an outage of $(k,m) \in \mathcal{E}$ and let $\mathrm{PTDF}^{\neg
    km}$ be the $(\abs{\mathcal{E}}-1) \times \abs{\mathcal{N}}$ PTDF matrix of the
    network with $(k,m)$ removed. Additionally, let $p_{\mathrm{f}}^{\neg km}\in
    \mathbb{R}^{\abs{\mathcal{E}}-1}$ be the vector of power flows across all of
    the remaining (still intact) branches $\mathcal{E}^{\neg km}\doteq \mathcal{E}
    \setminus \{(k,m) \}$. The power flow at $(l,t) \in \mathcal{E}^{\neg km}$
    reads
    \begin{equation}
        \label{eq:n1_case}p_{\mathrm{f},lt}^{\neg km}= \mathrm{PTDF}_{lt}^{\neg
        km}\left( \mathbf{C}_{\mathrm{G}}p_{\mathrm{G}}+ \mathbf{C}_{\mathrm{R}}p
        _{\mathrm{R}}- \mathbf{C}_{\mathrm{D}}p_{\mathrm{D}}\right ),
    \end{equation}
    where $p_{\mathrm{f},lt}^{\neg km}$ and $\mathrm{PTDF}_{lt}^{\neg km}$ are
    the $(l,t)$-th entry of $p_{\mathrm{f}}^{\neg km}$ and $(l,t)$-th row of
    $\mathrm{PTDF}^{\neg km}$, respectively.

    To find an (N-1)-secure operating point efficiently (i.e., without including
    $\mathcal{O}(\abs{\mathcal{E}}^{2})$ constraints for all possible branches
    and outage cases) we consider ranking the outages by severity, such that the
    most severe cases can be handled first. This leads to critical branch--critical
    outage (CBCO) combinations.

    \begin{definition}[CBCO and CBCO Analysis]
        \label{def:cbco} Consider an outage case $(k,m) \in \mathcal{E}$ as critical
        if it leads to a violated power flow on at least one branch in $\mathcal{E}
        ^{\neg km}$. Let $\mathcal{V}^{\neg km}\subseteq \mathcal{E}^{\neg km}$
        denote the set of all violated (i.e., critical) branches in case of a
        critical outage at $(k,m)$. Further, let
        $v^{\neg km}\in \mathbb{R}^{\abs{\mathcal{V}^{\neg km}}}$ store the
        magnitudes of the violations corresponding to each branch in
        $\mathcal{V}^{\neg km}$. A CBCO combination is then defined as a tuple $(
        (k,m), \mathcal{V}^{\neg km})$. We call the procedure of finding the set
        of all critical CBCO combinations the CBCO analysis\footnote{For the
        sake of completeness, inside the CBCO analysis we also check the base case
        with no branch outages.}.
    \end{definition}

    Using the CBCO analysis, one can introduce Algorithm~\ref{alg:n1_redispatch}
    based on the constraint generation method, where constraints of the form $\lvert
    p_{\mathrm{f},lt}^{\neg km}\rvert\leq p_{\mathrm{f},lt}^{\mathrm{max}}$ (cf.~\eqref{eq:n1_case})
    are repeatedly appended to \eqref{eq:redispatch} until its solution results
    in an operating point with no critical outages.
    \begin{algorithm}
        [t]
        \caption{Iterative (N-1)-secure deterministic redispatch}
        \label{alg:n1_redispatch}
        \begin{algorithmic}
            \REQUIRE Demands $p_{\mathrm{D}}$, feed-ins
            $(p_{\mathrm{G}}^{\bullet}, p_{\mathrm{R}}^{\bullet})$ after
            \newline
            market clearing \STATE $i\gets 0$, \, $\mathrm{Opt}_{\mathrm{RD}}\gets
            \text{optimization problem \eqref{eq:redispatch}}$ \STATE $(p_{\mathrm{G}}
            ^{+}, p_{\mathrm{G}}^{-}, p_{\mathrm{R}}^{-}) \gets \mathrm{solve}(\mathrm{Opt}
            _{\mathrm{RD}})$ \STATE $\mathrm{CBCO}_{i}\gets \mathrm{CBCO\_analysis}
            (p_{\mathrm{G}}^{\bullet}+p_{\mathrm{G}}^{+}-p_{\mathrm{G}}^{-}, p_{\mathrm{R}}
            ^{\bullet}-p_{\mathrm{R}}^{-})$ \WHILE {$\mathrm{CBCO}_{i}\neq \emptyset$}
            \STATE $i \gets i+ 1$ \STATE $( (k,m), \mathcal{V}^{\neg km}) \gets \mathrm{max\_violation\_outage}
            (\mathrm{CBCO}_{i-1})$ \FORALL{$(l,t) \in \mathcal{V}^{\neg km}$}
            \STATE add constraints \eqref{eq:n1_case} and $\abs{p_{\mathrm{f},lt}^{\neg km}}
            \leq p_{\mathrm{f},lt}^{\mathrm{max}}$ to $\mathrm{Opt}_{\mathrm{RD}}$
            \ENDFOR \STATE $(p_{\mathrm{G}}^{+}, p_{\mathrm{G}}^{-}, p_{\mathrm{R}}
            ^{-}) \gets \mathrm{solve}(\mathrm{Opt}_{\mathrm{RD}})$ \STATE $\mathrm{CBCO}
            _{i}\gets \mathrm{CBCO\_analysis}(p_{\mathrm{G}}^{\bullet}+p_{\mathrm{G}}
            ^{+}-p_{\mathrm{G}}^{-}, p_{\mathrm{R}}^{\bullet}-p_{\mathrm{R}}^{-})$
            \ENDWHILE
        \end{algorithmic}
    \end{algorithm}
    The function $\mathrm{max\_violation\_outage}$ searches for the CBCO combination
    $((k,m), \mathcal{V}^{\neg km})$ with the highest maximum violation $\max (v^{\neg
    km})$ among all of its critical branches. The search is performed within all
    of the CBCO combinations found by the CBCO analysis. This strategy resembles
    the cutting-plane method considered in \cite[Procedure~2.6]{bienstockChanceConstrainedOptimalPower2014}.

    In most applications, it is useful to consider the influence of uncertain
    variables of exogeneous nature on the parameters of the problem in \eqref{eq:redispatch}.
    These uncertainties may include but are not limited to: i) Demand-side
    quantities (e.g., simultanity of consumption); and ii) Quantities tied to
    forecast errors on the available renewable power, such as solar irradiance,
    wind speed, ambient temperature, etc.

    \subsection{Polynomial Chaos Expansions}
    In this section, we do not intend to provide an exhaustive overview of PCEs;
    for a more comprehensive dicussion we refer the interested reader to
    \cite{sullivanIntroductionUncertaintyQuantification2015} and references
    therein.

    To represent the dependencies on realizations of known (or modeled)
    uncertainties $\omega$, we consider a probability space $(\Omega, \mathcal{A}
    , \mathbb{P})$. This is a triplet, comprising: i) the set $\Omega$ that contains
    all possible uncertainty realizations (outcomes); ii) the set $\mathcal{A}$ which
    is a $\sigma$-algebra over $\Omega$ that describes all possible events associated
    with the uncertainties; and iii) the probability measure $\mathbb{P}:\mathcal{A}
    \to [0, 1]$ according to which the uncertainties are sampled, i.e.,
    $\omega \sim \mathbb{P}$. We define a real-valued random variable $X$ as a measurable
    function mapping an outcome $\omega$ to a realization $X(\omega) \in \mathbb{R}$.
    We define the expectation of $X$ as
    $\mathbb{E}[X] \doteq \int_{\tau \in \Omega}X(\tau) \mathbb{P}(\dd \tau )$
    and the variance of $X$ as
    $\mathbb{V}[ X] \doteq \mathbb{E}[( X-\mathbb{E}[X])^{2}]$. If $\mathbb{V}[X]
    < \infty$, we use the short-hand $X \in L^{2}(\mathbb{P}; \mathbb{R})$ to
    indicate that $X$ belongs to the space of square-integrable random variables
    with realizations in $\mathbb{R}$. Since $L^{2}(\mathbb{P}; \mathbb{R})$ is a
    Hilbert space, it can be endowed with an inner product
    \[
        \langle X, Y \rangle_{L^{2}}\doteq \mathbb{E}[XY] = \int_{\tau \in
        \Omega}X( \tau)Y(\tau) \mathbb{P}(\dd \tau).
    \]

    Considering a scalar uncertainty $\omega$ with realizations in $\mathbb{R}$,
    one can construct a series of polynomials
    $\left\{\phi^{i}\right\}_{i \in \mathbb{N}_0}$ in the indeterminate $\omega$
    of increasing degree $i$ such that these polynomials are pairwise
    orthonormal, i.e.,
    \[
        \langle \phi^{i}, \phi^{j}\rangle_{L^{2}}= \delta_{ij},
    \]
    where $\delta_{ij}$ is the Kronecker delta. Depending on the distribution
    $\mathbb{P}$, the polynomials can be adjusted to preserve orthogonality. For
    instance, there exist families of orthogonal polynomials (normalized by
    $\langle \phi^{i}, \phi^{i}\rangle_{L^{2}}$), such as the Legendre
    polynomials for uniform $\mathbb{P}$, Hermite polynomials for Gaussian $\mathbb{P}$,
    etc. \cite{xiuWienerAskeyPolynomialChaos2002}.

    To generalize the orthogonal polynomials to the (more practically relevant) case
    of a vector of $n_{\omega}\in \mathbb{N}$ uncertainties $\omega \doteq \mqty
    [\omega_{1}\ldots \omega_{n_\omega}]^{\top}$, we consider individual
    distributions $\omega_{i}\sim \mathbb{P}_{i}$ for each
    $i \in \mathbb{I}_{n_\omega}$. Assuming that all uncertainties are mutually independent,
    the joint distribution is given as a product measure
    $\mathbb{P}= \bigotimes_{i = 1}^{n_\omega}\mathbb{P}_{i}$. Hence, the
    multivariate polynomials can be assembled as products
    $\bm{\phi}^{\alpha}= \prod_{j=1}^{n_\omega}\phi^{\alpha_j}$, where $\alpha =
    \mqty[\alpha_{1}\ldots \alpha_{n_\omega}]^{\top}\in \mathbb{N}_{0}^{n_\omega}$
    is a multi-index, i.e., each $\alpha_{j}$ specifies the degree of the
    polynomial in the $j$-th dimension. Now, different families of orthogonal
    polynomials can be selected for each dimension $j$ according to
    $\mathbb{P}_{j}$.

    Because the polynomials depend on $\omega$, one can see them as random variables
    in $L^{2}( \mathbb{P}; \mathbb{R})$. The central idea of PCEs is to
    represent any $X \in L^{2}(\mathbb{P}; \mathbb{R})$ as a linear combination
    \begin{equation*}
        X = \sum_{\abs{\alpha} = 0}^{\infty}\pcec{x}{\alpha}\bm{\phi}^{\alpha},~\text{where}
        ~ \pcec{x}{\alpha}\in \mathbb{R},~\alpha \in \mathbb{N}_{0}^{n_\omega}
    \end{equation*}
    of orthogonal polynomials of increasing total degree
    $\abs{\alpha}\doteq \sum_{j=1}^{n_\omega}\alpha_{j}$. Although the convergence
    is guaranteed for any random variable in $L^{2}(\mathbb{P}; \mathbb{R})$ \cite{sullivanIntroductionUncertaintyQuantification2015},
    the infinite sum above might not be suitable in practice. Hence, it is often
    beneficial to use a finite-dimensional basis approximation
    \begin{equation}
        \label{eq:exact_pce}X \approx \Tilde{X}= \sum_{\alpha \in M}\pcec{x}{\alpha}
        \bm{\phi}^{\alpha}.
    \end{equation}
    on a (finite) subset $M$ of multi-indices. Here, we denote the vector of all
    PCE coefficients as
    $\pcec{x}{M}\doteq [\pcec{x}{\alpha}]_{\alpha \in M}\in \mathbb{R}^{1 \times
    \abs{M}}$. If $X \in L^{2}(\mathbb{P}; \mathbb{R}^{n_x})$ is vector-valued of
    dimension $n_{x}\in \mathbb{N}$, then its PCE coefficients are also vectors,
    i.e., $\pcec{x}{\alpha}\in \mathbb{R}^{n_x}$ for each $\alpha \in M$. In
    this case, $\pcec{x}{M}\in \mathbb{R}^{n_x \times \abs{M}}$ is a matrix. The
    approximation error in \eqref{eq:exact_pce} depends on the choice of $M$ \cite{muhlpfordtCommentsTruncationErrors2018}.

    For the remainder of the paper, we indicate random variables by writing them
    in regular upper-case font (e.g., $X$) and denote their PCE coefficients in
    lower-case sans serif, with the corresponding (sets of) multi-indices written
    in superscripts (e.g., $\pcec{x}{\alpha}$ or $\pcec{x}{M}$).

    The PCE representation \eqref{eq:exact_pce} allows one to express the
    expectation and variance of $\tilde{X}\in L^{2}(\mathbb{P}; \mathbb{R})$ via
    \begin{equation}
        \label{eq:pce_moments}\mathbb{E}[\tilde{X}]= \pcec{x}{\mathbf{0}},~\text{and}
        ~\mathbb{V}[\tilde{X}]= \sum\nolimits_{\alpha \in M\setminus\{\mathbf{0}\}}
        (\pcec{x}{\alpha})^{2},
    \end{equation}
    where $\mathbf{0}$ is the multi-index with all entries equal to zero. This is
    particularly useful as it allows one to analyze the moments of the distribution
    of a random variable in $L^{2}(\mathbb{P}; \mathbb{R})$ by only using the coefficients
    of its (approximated) PCE representation.

    \subsection{Stochastic Redispatch}
    In general, $L^{2}(\mathbb{P}; \mathbb{R})$ is infinite-dimensional. However,
    if all parameters of \eqref{eq:redispatch} that depend on $\omega$ admit a finite-dimensional
    representation \eqref{eq:exact_pce} in a common orthogonal polynomial basis,
    we can restrict ourselves to the finite-dimensional subspace of $L^{2}(\mathbb{P}
    ; \mathbb{R})$, where each random variable can be described using finitely
    many PCE coefficients. This motivates the following standing assumptions:
    \begin{itemize}
        \item The uncertainty stems from the RES generation forecasts;

        \item Market clearing utilizes all of the available renewable generation
            without curtailment $\mathbb{P}$-almost surely; \emph{and},

        \item The RES generation forercasts are approximated via
            \begin{equation}
                \label{eq:uncertainty_source}P_{\mathrm{R}}^{\bullet}= \sum_{\alpha
                \in M}\mathsf{p}_{\mathrm{R}}^{\bullet, \alpha}\bm{\phi}^{\alpha}
                .
            \end{equation}
            indexed by a (finite) set of multi-indices $M$.
    \end{itemize}

    Relying on these assumptions and following \cite{muhlpfordtGeneralizedFrameworkChanceconstrained2018},
    we formulate the stochastic version of \eqref{eq:redispatch} by exchanging the
    previously deterministic $p_{\mathrm{R}}^{\bullet}\in \mathbb{R}^{\abs{\mathcal{R}}}$
    in \eqref{eq:redispatch} for its PCE representation \eqref{eq:uncertainty_source}.
    We apply the same substitution to obtain $P_{\mathrm{G}}^{\bullet}$, $P_{\mathrm{G}}
    ^{+}$, $P_{\mathrm{G}}^{-}$, $P_{\mathrm{R}}^{-}$, and $P_{\mathrm{f}}$. The
    resulting optimization problem reads
    \begin{flalign}
        \label{eq:stoch_redispatch} & \underset{\mathsf{p}_{\mathrm{G}}^{+, M}, \mathsf{p}_{\mathrm{G}}^{-, M}, \mathsf{p}_{\mathrm{R}}^{-, M}}{\text{minimize}}~\expv{f(P_\mathrm{G}^+, P_\mathrm{G}^-, P_\mathrm{R}^-)}                                                                                                                                                                                                                                                                                                                                                                                                                                                                                                                                                                                                                                                                                                                                                                                                                                                                                                                                                                                                                                                                                                                                                                                                                                                                                                                                                                                                                                                                                & \nonumber \\
                                    & \text{subject to:}                                                                                                                                                                                                                                                                                                                                                                                                                                                                                                                                                                                                                                                                                                                                                                                                                                                                                                                                                                                                                                                                                                                                                                                                                                                                                                                                                                                                                                                                                                                                                                                                                                                 &           \\
                                    & \begin{alignedat}{2} & \textstyle\sum\limits_{i \in \mathcal{G}}\pcec{p}{\alpha}_{\mathrm{G},i}+ \textstyle\sum\limits_{i \in \mathcal{R}}\pcec{p}{\alpha}_{\mathrm{R},i}- \textstyle\sum\limits_{i \in \mathcal{D}}\pcec{p}{\alpha}_{\mathrm{D},i}= 0, & & \,\forall \,\alpha \in M, \\ & \pcec{p}{\alpha}_{\mathrm{f}}=\mathrm{PTDF}\left( \mathbf{C}_{\mathrm{G}}\pcec{p}{\alpha}_{\mathrm{G}}+ \mathbf{C}_{\mathrm{R}}\pcec{p}{\alpha}_{\mathrm{R}}- \mathbf{C}_{\mathrm{D}}\pcec{p}{\alpha}_{\mathrm{D}}\right), & & \,\forall \,\alpha \in M, \\ & \pcec{p}{\alpha}_{\mathrm{G}}= \mathsf{p}_{\mathrm{G}}^{\bullet, \alpha}+ \mathsf{p}_{\mathrm{G}}^{+, \alpha}- \mathsf{p}_{\mathrm{G}}^{-, \alpha}, ~~\pcec{p}{\alpha}_{\mathrm{R}}= \mathsf{p}_{\mathrm{R}}^{\bullet, \alpha}- \mathsf{p}_{\mathrm{R}}^{-, \alpha}, & & \,\forall \,\alpha \in M, \\ & 0 \leq \mathsf{p}_{\mathrm{G}}^{+, \mathbf{0}}\leq p_{\mathrm{G},i}^{+,\max}, & & \,\forall \, i \in \mathcal{G}, \\ & 0 \leq \mathsf{p}_{\mathrm{G}}^{+, \mathbf{0}}\pm \lambda(\varepsilon) \sqrt{\textstyle\sum\limits_{\alpha \in M\setminus\{\mathbf{0}\}}(\mathsf{p}_{\mathrm{G}}^{+, \alpha})^{2}}\leq p_{\mathrm{G},i}^{+,\max}, & & \,\forall \, i \in \mathcal{G}, \\ & \vdots & & \,\vdots \\ & \abs{\pcec{p}{\mathbf{0}}_{\mathrm{f},km}}\leq p_{\mathrm{f},km}^{\max}, & & \,\forall \, (k,m) \in \mathcal{E}, \\ & \abs{\pcec{p}{\mathbf{0}}_{\mathrm{f},km}\pm \lambda(\varepsilon) \sqrt{\textstyle\sum\limits_{\alpha \in M\setminus\{\mathbf{0}\}}(\pcec{p}{\alpha}_{\mathrm{f},km})^{2}}}\leq p_{\mathrm{f},km}^{\max}, & & \,\forall \, (k,m) \in \mathcal{E}.\end{alignedat} & \nonumber
    \end{flalign}
    Notice that we use a risk-neutral setting \cite{roaldPowerSystemsOptimization2023},
    i.e., we minimize the expected value of the cost function. This expectation is
    expanded by substituting the PCE representations of $P_{\mathrm{G}}^{+}$,
    $P_{\mathrm{G}}^{-}$, and $P_{\mathrm{R}}^{-}$ into \eqref{eq:objective} and
    making use of the relation $\expv{X^2}= \expv{X}^{2}+ \varv{X}$ for the
    quadratic terms. Note also that for the vector of demands $P_{\mathrm{D}}$ we
    have that $\pcec{p}{\alpha}_{\mathrm{D}}= p_{\mathrm{D}}$ if $\alpha = \mathbf{0}$
    and $\pcec{p}{\alpha}_{\mathrm{D}}= 0$ otherwise, i.e., $P_{\mathrm{D}}$
    remains deterministic.

    In \eqref{eq:stoch_redispatch}, we rewrite the equality constraints in terms
    of the PCE coefficients, whereas the inequalities are formulated as chance-constraints.
    Here, we set $\lambda(\varepsilon) \doteq \sqrt{1/\varepsilon}$ according to
    the Chebyshev inequality, which is valid for any $L^{2}$-distribution.

    The decision variables of \eqref{eq:stoch_redispatch} are the matrices $\mathsf{p}
    _{\mathrm{G}}^{+, M}, \mathsf{p}_{\mathrm{G}}^{-, M}\in \mathbb{R}^{\abs{\mathcal{G}}
    \times \abs{M}}$ and $\mathsf{p}_{\mathrm{R}}^{-, M}\in \mathbb{R}^{\abs{\mathcal{R}}
    \times \abs{M}}$ of PCE coefficients for the adjustments to the power of conventional
    and RES generators, respectively. Suppose that $(\mathsf{p}_{\mathrm{G}}^{+,
    M})^{\star}$ $(\mathsf{p}_{\mathrm{G}}^{-, M})^{\star}$, and
    $(\mathsf{p}_{\mathrm{R}}^{-, M})^{\star}$ are the corresponding solutions
    to \eqref{eq:stoch_redispatch}. One can then recover the optimal random-variable
    decisions by substitutions
    \begin{equation}
        \label{eq:solution_recovery}(P_{\mathrm{G}}^{s})^{\star}= \textstyle\sum
        \limits_{\alpha \in M}(\mathsf{p}_{\mathrm{G}}^{s, \alpha})^{\star}\bm{\phi}
        ^{\alpha}~\text{and}~(P_{\mathrm{R}}^{-})^{\star}= \textstyle\sum\limits_{\alpha
        \in M}(\mathsf{p}_{\mathrm{R}}^{-, \alpha})^{\star}\bm{\phi}^{\alpha}
    \end{equation}
    into the respective exact PCE representations with $s \in \{+,-\}$. The recovered
    solution depends on the uncertainties $\omega$ and this dependency is computable
    provided that the polynomials $\bm{\phi}^{\alpha}$ in \eqref{eq:uncertainty_source}
    are known \cite{muhlpfordtGeneralizedFrameworkChanceconstrained2018}. This allows
    us to adjust the power generation based on the realization of the uncertainties,
    unlike having a constant solution as in \eqref{eq:redispatch}.

    Note that every chance constraint is convex since it draws a bicone in the space
    of PCE coefficients of its corresponding variable. Therefore, due to the quadratic
    cost functions \eqref{eq:objective}, we can classify \eqref{eq:stoch_redispatch}
    as a second-order cone program, which can be solved efficiently \cite{loboApplicationsSecondorderCone1998}.

    \section{Proposed Approach For (N-1)-Secure Stochastic OPF}
    \label{sec:main_results} To this end, we propose a method to augment \eqref{eq:stoch_redispatch}
    with a procedure akin to Algorithm~\ref{alg:n1_redispatch} for deterministic
    redispatch \eqref{eq:redispatch}. We also suggest a technique to construct polynomial
    basis for generation forecasts of wind and solar-based RES with modeling of correlations
    between different units.
    \subsection{PCE-based (N-1)-Secure Redispatch}
    To begin assessing the (N-1)-security of the solution to \eqref{eq:stoch_redispatch},
    we can start by obtaining the nodal net power injections
    \[
        (P_{\mathrm{net}})^{\star}= \mathbf{C}_{\mathrm{G}}(P_{\mathrm{G}})^{\star}
        + \mathbf{C}_{\mathrm{R}}(P_{\mathrm{R}})^{\star}- \mathbf{C}_{\mathrm{D}}
        p_{\mathrm{D}}
    \]
    where $(P_{\mathrm{G}})^{\star}$ and $(P_{\mathrm{R}})^{\star}$ are obtained
    using \eqref{eq:solution_recovery} and the random-variable counterparts of the
    power adjustment constraints \eqref{eq:red_pg_update}. Using the net injections,
    we can obtain the power flows by multiplication with different PTDF matrices,
    depending on the outage case of interest, i.e.,
    \[
        \left\{
        \begin{array}{ll}
            (P_{\mathrm{f}})^{\star} = \mathrm{PTDF} (P_{\mathrm{net}})^{\star},                     & \text{for no outage, or}                    \\
            (P_{\mathrm{f}}^{\neg km})^{\star} = \mathrm{PTDF}^{\neg km} (P_{\mathrm{net}})^{\star}, & \text{for outage of}~(k,m) \in \mathcal{E}.
        \end{array}
        \right.
    \]
    The resulting power flows are random-variable vectors
    $(P_{\mathrm{f}})^{\star}$ and $(P_{\mathrm{f}}^{\neg km})^{\star}$. Hence,
    the previously mentioned Definition~\ref{def:cbco} of CBCOs needs to be
    adapted as it is only suitable for deterministic quantities. To classify outages
    as critical and to be able to compare outages in terms of their severity, we
    use the probabilistic violation criterion we propose below.

    \begin{definition}[$\epsilon$-CBCO and $\epsilon$-CBCO Analysis]
        \label{def:eps_cbco} Let $\varepsilon \in (0,1)$ be a user-defined failure
        rate. Consider an outage of $(k,m)\in \mathcal{E}$ to be $\varepsilon$-critical
        if there exists at least one intact branch
        $(l,t) \in \mathcal{E}^{\neg km}$ such that
        \begin{equation}
            \label{eq:stoch_violation}\mathbb{P}\left(\left\{\omega \in \Omega \mid
            \abs{(P_{\mathrm{f},lt}^{\neg km})^{\star}(\omega)}\geq p_{\mathrm{f},lt}
            ^{\max}\right\}\right) > \varepsilon.
        \end{equation}
        Let
        $\mathcal{V}^{\neg km}(\varepsilon ) \subseteq \mathcal{E}^{\neg km}$
        denote the set of branches $(l,t)$ that meet the above criterion. To quantify
        the severity of the individual branch violations, we introduce the
        vector
        $v^{\neg km}(\varepsilon ) \in \mathbb{R}^{\abs{\mathcal{V}^{\neg km}(\varepsilon )}}$
        that stores the expected violations
        \begin{equation}
            \label{eq:mean_violation}v^{\neg km}_{lt}(\varepsilon ) \doteq \expv{\abs{(P_{\mathrm{f},lt}^{\neg km})^{\star}} - p_{\mathrm{f},lt}^{\max}}
        \end{equation}
        for each $(l,t) \in \mathcal{V}^{\neg km}(\varepsilon )$.

        An $\varepsilon$-CBCO combination is then defined as a tuple
        $( (k,m), \mathcal{V}^{\neg km}(\varepsilon ) )$. We call the procedure of
        finding the set of all critical $\varepsilon$-CBCO combinations the $\varepsilon$-CBCO
        analysis.
    \end{definition}
    In practice, the cumulative density function of $(P_{\mathrm{f},lt}^{\neg km}
    )^{\star}$ might be difficult to find analytically. Hence, we use a MC method
    to check the violation criterion \eqref{eq:stoch_violation}. Let $\{\omega^{(i)}
    , \ldots, \omega^{(N_\mathrm{samples})}\}$ be the $N_{\mathrm{samples}}$ sample
    realizations of the known uncertainties $\omega$. The probability of
    violation can be approximated by evaluating
    \begin{subequations}
        \label{eq:stoch_violation_mc}
        \begin{equation}
            N_{\mathrm{samples}}^{\mathrm{violated}}\doteq \abs{\{i \in \mathbb{I}_{N_\mathrm{samples}} \mid \abs{(P_{\mathrm{f},lt}^{\neg km})^{\star}(\omega^{(i)})}\geq p_{\mathrm{f},lt} ^{\max}\}}
            ,
        \end{equation}
        i.e., counting the number of samples resulting in violation. Then, \eqref{eq:stoch_violation}
        can be reformulated as
        \begin{equation}
            N_{\mathrm{samples}}^{\mathrm{violated}}/ N_{\mathrm{samples}}> \varepsilon
            .
        \end{equation}
        Similarly, \eqref{eq:mean_violation} can be approximated by
        \begin{equation}
            v^{\neg km}_{lt}(\varepsilon ) \approx \frac{1}{N_{\mathrm{samples}}}
            \textstyle \sum\limits_{i=1}^{N_\mathrm{samples}}\abs{(P_{\mathrm{f},lt}^{\neg km})^{\star}(\omega^{(i)})}
            - p_{\mathrm{f},lt}^{\max}.
        \end{equation}
    \end{subequations}

    The stochastic version of Algorithm~\ref{alg:n1_redispatch} is formulated in
    Algorithm~\ref{alg:n1_stoch_redispatch} by replacing the deterministic redispatch
    \eqref{eq:redispatch} with its stochastic formulation \eqref{eq:stoch_redispatch}
    and by using $\varepsilon$-CBCO analysis as per Definition~\ref{def:eps_cbco}
    instead of the deterministic CBCO analysis from Definition~\ref{def:cbco}.
    In the following, we discuss the estimation of the PCE coefficients $\mathsf{p}
    _{\mathrm{R}}^{\bullet, M}\in \mathbb{R}^{\abs{\mathcal{R}} \times \abs{M}}$
    and $\mathsf{p}_{\mathrm{G}}^{\bullet, M}\in \mathbb{R}^{\abs{\mathcal{G}} \times
    \abs{M}}$ that specify the market clearing scenarios for our stochastic setting
    in Algorithm~\ref{alg:n1_stoch_redispatch}.

    \begin{algorithm}
        [t]
        \caption{Iterative (N-1)-secure stochastic redispatch}
        \label{alg:n1_stoch_redispatch}
        \begin{algorithmic}
            \REQUIRE Demands $p_{\mathrm{D}}\in \mathbb{R}^{\abs{\mathcal{D}}}$;
            PCE coefficients
            $\mathsf{p}_{\mathrm{R}}^{\bullet, M}\in \mathbb{R}^{\abs{\mathcal{R}}
            \times \abs{M}}$
            and
            $\mathsf{p}_{\mathrm{G}}^{\bullet, M}\in \mathbb{R}^{\abs{\mathcal{G}}
            \times \abs{M}}$
            for RES generation forecasts and conventional generation market
            clearing, respectively; failure rate $\varepsilon \in (0,1)$.
            \STATE $i\gets 0$, \,
            $\mathrm{Opt}_{\mathrm{RD}}\gets \text{optimization problem \eqref{eq:stoch_redispatch}}$
            \STATE
            $(\mathsf{p}_{\mathrm{G}}^{+,M}, \mathsf{p}_{\mathrm{G}}^{-,M}, \mathsf{p}
            _{\mathrm{R}}^{-,M}) \gets \mathrm{solve}(\mathrm{Opt}_{\mathrm{RD}})$
            \STATE Obtain $(P_{\mathrm{G}}, P_{\mathrm{R}})$ using solution
            recovery \eqref{eq:solution_recovery}
            \STATE $\varepsilon\text{-}\mathrm{CBCO}_{i}\gets \varepsilon\text{-}
            \mathrm{CBCO\_analysis}(P_{\mathrm{G}}, P_{\mathrm{R}})$
            \WHILE {$\varepsilon\text{-}\mathrm{CBCO}_{i}\neq \emptyset$}
            \STATE $i \gets i+ 1$
            \STATE $( (k,m), \mathcal{V}^{\neg km}) \gets \mathrm{max\_mean\_violation}
            ( \varepsilon\text{-}\mathrm{CBCO}_{i-1})$
            \FORALL{$(l,t) \in \mathcal{V}^{\neg km}$}
            \STATE add chance constraints for
            $\abs{P_{\mathrm{f},lt}^{\neg km}}$ to $\mathrm{Opt}_{\mathrm{RD}}$
            \ENDFOR
            \STATE
            $(\mathsf{p}_{\mathrm{G}}^{+,M}, \mathsf{p}_{\mathrm{G}}^{-,M}, \mathsf{p}
            _{\mathrm{R}}^{-,M}) \gets \mathrm{solve}(\mathrm{Opt}_{\mathrm{RD}})$
            \STATE Obtain $(P_{\mathrm{G}}, P_{\mathrm{R}})$ using solution
            recovery \eqref{eq:solution_recovery}
            \STATE $\varepsilon\text{-}\mathrm{CBCO}_{i}\gets \varepsilon\text{-}
            \mathrm{CBCO\_analysis}(P_{\mathrm{G}}, P_{\mathrm{R}})$
            \ENDWHILE
        \end{algorithmic}
    \end{algorithm}

    \subsection{Polynomial Basis for RES Generation Forecasts}
    To complete the formulation of the PCE-based stochastic redispatch with (N-1)-security,
    we first specify the polynomial basis used in \eqref{eq:uncertainty_source}
    such that the wind and solar generation forecasts \eqref{eq:uncertainty_source}
    admit realistic distributions.

    Typically, forecasts are provided based on available statistical data
    obtained from a history of observations. As a starting point, we assume that
    the data available about $P^{\bullet}_{\mathrm{R},j}$ consists of a bundle
    of the following scalar values:
    \begin{itemize}
        \item Mean $\mu_{\mathrm{R},j}$ s.t.
            $\expv{P^{\bullet}_{\mathrm{R},j}}= \mu_{\mathrm{R},j}$;

        \item $5\%$-quantile $q^{5}_{\mathrm{R},j}$ s.t.
            $\mathbb{P}\{\omega \mid P^{\bullet}_{\mathrm{R},j}(\omega )\leq q^{5}
            _{\mathrm{R},j}\} = 0.05$;

        \item $95\%$-quantile $q^{95}_{\mathrm{R},j}$ s.t.
            $\mathbb{P}\{ \omega \mid P^{\bullet}_{\mathrm{R},j}(\omega ) \leq q^{95}
            _{\mathrm{R},j}\} = 0.95$;

        \item and values $a_{\mathrm{R},j},b_{\mathrm{R},j}$ specifying the support
            interval $[a_{\mathrm{R},j},b_{\mathrm{R},j}]$ of the distribution of
            $P^{\bullet}_{\mathrm{R},j}$.
    \end{itemize}
    In addition, assume that all entries in $P^{\bullet}_{\mathrm{R}}$ are
    mutually independent.

    \begin{remark}[Forecast dependencies]
        \label{rem:dependencies} While the assumption of independency is somewhat
        restrictive, there exist some remedies. For instance, given the marginal
        distributions and the correlation matrix of a vector-valued random
        variable, a corresponding vector-valued PCE approximation can be obtained
        using the approaches discussed in \cite{jakemanPolynomialChaosExpansions2019}
        and based on a sufficiently large number of sample realizations.

        Due to the practical relevance of dependencies, in this paper, we show an
        ad-hoc approach to embed correlations into the forecast PCE series after
        first computing the coefficients for the independent case.
    \end{remark}

    To model the RES power generation forecasts, we select beta distributions
    $\mathrm{Beta}(\alpha_{j}, \beta_{j})$, with parameters $\alpha_{j}, \beta_{j}
    > 0, \forall \, j \in \mathcal{R}$. This choice is motivated by the fact
    that beta distributions: i) Have bounded support (unlike, e.g., Gaussians);
    ii) Allow for adjustable skewness; and iii) Have finite variance (i.e.,
    belong to the $L^{2}$ space), and hence admit a corresponding series of
    orthogonal polynomials. For example usage we refer to \cite{fernandez-jimenezShorttermProbabilisticForecasting2023}.

    Parameters $\alpha_{j}$ and $\beta_{j}$ of the individual (marginal) beta-distributions
    of the RES generation forecasts are found by fitting them to the available data
    via solving
    \begin{equation}
        \label{eq:beta_estimation}\underset{\alpha \in [0, \Bar{\alpha}], \, \beta \in [0, \Bar{\beta}]}
        {\text{minimize}}~~ f_{\mathrm{error},j}(\alpha, \beta)
    \end{equation}
    for each $j \in \mathcal{R}$. The parameters are box-constrained with upper
    bounds $\Bar{\alpha}, \Bar{\beta}\in \mathbb{R}$, whereas the objective $f_{\mathrm{error}}$
    here is defined as the sum of the squared mismatches in terms of the quantile
    and expected value data $(\mu_{\mathrm{R},j}, q^{5}_{\mathrm{R},j}, q^{95}_{\mathrm{R},j}
    )$. Using the inverse cumulative distribution function
    $F_{\mathrm{beta},\alpha,\beta}^{-1}: [ 0,1] \to \mathbb{R}$ of
    $\mathrm{Beta}(\alpha,\beta)$, it can be formulated as
    \begin{multline*}
        f_{\mathrm{error},j}(\alpha, \beta) \doteq (F_{\mathrm{beta},\alpha,\beta}
        ^{-1}(0.95)-q^{5}_{\mathrm{R},j})^{2}+ \\
        (F_{\mathrm{beta},\alpha,\beta}^{-1}(0.05)-q^{95}_{\mathrm{R},j})^{2}+ \left
        (\frac{\alpha}{\alpha +\beta}- \mu_{\mathrm{R},j}\right)^{2},~~\forall \,
        j \in \mathcal{R}.
    \end{multline*}

    Since a beta distribution has a well-defined variance, using the Jacobi series
    of polynomials, any beta-distributed random variable can be modeled by a two-term
    PCE series. Hence, for the PCE representation~\eqref{eq:uncertainty_source},
    we choose an affine structure
    \begin{equation}
        \label{eq:affine_pce}M \doteq \{\alpha \in \mathbb{N}_{0}^{\abs{\mathcal{R}}}
        \mid \abs{\alpha}\le 1 \}
    \end{equation}
    with $\abs{M}= \abs{\mathcal{R}}+ 1$ terms, consisting of one term corresponding
    to the mean ($\mathbf{0}$-th PCE coefficient multiplied with polynomial
    $\bm{\phi}^{\mathbf{0}}= 1$) and $\abs{\mathcal{R}}$ terms corresponding to the
    first-order Jacobi polynomials representing forecast uncertainties for each RES.

    Going back to Remark~\ref{rem:dependencies}, we consider the case in which a
    $\abs{\mathcal{R}}\times \abs{\mathcal{R}}$ correlation matrix $\mathbf{E}$
    is specified for the RES forecasts. To include this information into \eqref{eq:uncertainty_source},
    one can compute the entries of the desired covariance matrix $\bm{\Sigma}\in
    \mathbb{R}^{\abs{\mathcal{R}}\times \abs{\mathcal{R}}}$ using the formula $\sigma
    (\alpha, \beta) \doteq \sqrt{\alpha \beta /[(\alpha +\beta )^{2}(\alpha +\beta
    +1)]}$ to calculate the standard deviation of $\mathrm{Beta}(\alpha, \beta)$.
    The correlation-adjusted PCE coefficients can then be obtained by Cholesky
    decomposition of $\bm{\Sigma}$. We present the entire procedure for the estimation
    of the RES forecast PCE coefficients $\mathsf{p}_{\mathrm{R}}^{\bullet, M}$ in
    Algorithm~\ref{alg:pg_pcec}. To obtain the full probabilistic description of
    the market clearing, we pick $\mathsf{p}_{\mathrm{G}}^{\bullet, M}$ from a suitable
    subspace, i.e., such that the power conservation \eqref{eq:conservation} is satisfied
    for ($\mathbb{P}$-almost) all realizations $P_{\mathrm{G}}^{\bullet}(\omega)
    = \sum_{\alpha \in M}\mathsf{p}_{\mathrm{G}}^{\bullet, \alpha}\bm{\phi}^{\alpha}
    (\omega)$ and $P_{\mathrm{R}}^{\bullet}(\omega) = \sum_{\alpha \in M}\mathsf{p}
    _{\mathrm{R}}^{\bullet, \alpha}\bm{\phi}^{\alpha}(\omega)$.

    \begin{algorithm}
        [t]
        \caption{PCE for RES generation forecasts from data}
        \label{alg:pg_pcec}
        \begin{algorithmic}
            \REQUIRE Statistical RES generation data
            \newline
            $(\mu_{\mathrm{R}}, q^{5}_{\mathrm{R}}, q^{95}_{\mathrm{R}}, a_{\mathrm{R}}
            , b_{\mathrm{R}})$, correlations
            $\mathbf{E}\in \mathbb{R}^{\abs{\mathcal{R}}\times \abs{\mathcal{R}}}$.
            \FORALL{$j \in \{1, \ldots, \abs{\mathcal{R}}\}$}
            \STATE
            $\mathsf{p}_{\mathrm{R},j}^{\bullet, \mathbf{0}}\gets \mu_{\mathrm{R},j}$
            \STATE $c_{\mathrm{R},j}\gets b_{\mathrm{R},j}- a_{\mathrm{R},j}$
            \STATE
            $(\mu_{\mathrm{R},j}, q^{5}_{\mathrm{R},j}, q^{95}_{\mathrm{R},j}) \gets
            [(\mu_{\mathrm{R},j}, q^{5}_{\mathrm{R},j}, q^{95}_{\mathrm{R},j})- a
            _{\mathrm{R},j}]/c_{\mathrm{R},j}$
            \STATE
            $(\alpha_{j}, \beta_{j}) \gets \text{solve~\eqref{eq:beta_estimation}~for~}
            (\mu_{\mathrm{R},j}, q^{5}_{\mathrm{R},j}, q^{95}_{\mathrm{R},j})$
            \STATE $\mathbb{P}_{j}\gets \mathrm{Beta}(\alpha_{j}, \beta_{j})$
            \STATE
            $\bm{\phi}^{\alpha}(\omega) \gets \left [\beta_{j}+ (\alpha_{j}+ \beta
            _{j})(\omega_{j}- 1) \right]$
            \FORALL{$i \in \mathcal{R}: i < j$}
            \STATE $\bm{\Sigma}_{ij}\gets \mathbf{E}_{ij}\sigma(\alpha_{i}, \beta
            _{i}) \sigma (\alpha_{j}, \beta_{j})$
            \ENDFOR
            \ENDFOR
            \STATE $\mathsf{p}_{\mathrm{R}}^{\bullet, M\setminus\{\mathbf{0}\}}\gets
            \mathrm{Cholesky}(\bm{\Sigma})$
        \end{algorithmic}
    \end{algorithm}

    \section{Numerical Example}
    In this section, we apply the stochastic (N-1)-secure redispatch algorithm to
    the IEEE 118-bus test case with $\mathcal{N}= \mathbb{I}_{118}$,
    $\abs{\mathcal{E}}= 186$, $\abs{\mathcal{G}}= 54$, and
    $\abs{\mathcal{D}}= 99$ \cite{christiePowerSystems1993}. The network data,
    including cost coefficients for the conventional generators, is taken from the
    Matpower database \cite{zimmermanMATPOWERSteadyStateOperations2011}. We
    modify the network by including $\abs{\mathcal{R}}= 24$ additional RES units,
    consisting of 7 wind-based and 17 solar-based power plants, located at buses
    $\{ 17,26,54,59,71,104,112\}$ and $\{1,$
    $4,10,45,47,56,61,69,77,80,87, 91,97,100,101,105,113\}$, respectively. The
    numerical example is coded in Julia using \texttt{PowerModels.jl} \cite{coffrinPowerModels2018}
    and \texttt{PolyChaos.jl} \cite{muhlpfordtPolyChaosjlJuliaPackage2020}, whereas
    the optimization problems \eqref{eq:redispatch} and \eqref{eq:stoch_redispatch}
    are solved using Mosek \cite{mosek}. The computations are done on a machine
    powered by Intel(R) Core\texttrademark i5-1335U with $16$~GB RAM.

    For each RES in $j \in \mathcal{R}$, besides its bus location, we specify
    the additional five parameters
    $(\mu_{\mathrm{R},j}, q^{5}_{\mathrm{R},j}, q^{95}_{\mathrm{R},j}, a_{\mathrm{R},j}
    , b_{\mathrm{R},j})$, defined previously. The largest units are the solar
    power plants at $54 \in \mathcal{N}$ and $71 \in \mathcal{N}$ with mean
    generation forecast values of $50$~MW and $40$~MW, respectively. Relying on
    the study in \cite{widenCorrelationsLargeScaleSolar2011}, we specify the the
    correlations between forecasts for any pair $i,j \in \mathcal{R}$ of RES as $\mathbf{E}
    _{ij}\doteq 0.85$ if both $i$ and $j$ are solar-powered,
    $\mathbf{E}_{ij}\doteq 0.6$ if if both $i$ and $j$ are wind-powered, and $\mathbf{E}
    _{ij}\doteq 0$ otherwise.

    We set all branch limits to $240$~MW for all branches, except $480$~MW for
    10 selected branches, $300$~MW for $\{(8,30), (26,30)\} \subset \mathcal{E}$,
    and $320$~MW for $\{(8,9), (9,10)\} \subset \mathcal{E}$. The maximum values
    for the redispatch adjustments are defined as follows. We allow the RES to
    be curtailed up to $20\%$ of their respective mean forecast values, whereas for
    the ramp-ups and ramp-downs of the conventional plants, the limits are set
    to $40\%$ and $6 0\%$ of their maximum power generations, respectively. For
    the chance-constraints in \eqref{eq:stoch_redispatch}, we set
    $\varepsilon = 0.05$.

    We compare the method introduced in Section~\ref{sec:main_results} with the
    MC approach. We start by obtaining the PCE coefficients of \eqref{eq:uncertainty_source}
    given the affine basis structure \eqref{eq:affine_pce} and estimation results
    for the parameters of the beta distributions based on \eqref{eq:beta_estimation}.
    Subsequently, we calculate the PCE coefficients corresponding to the
    conventional generation $P^{\bullet}_{\mathrm{G}}$. This allows us to generate
    $N_{\mathrm{samples}}= 10 000$ samples $\{\omega^{(i)}, \ldots , \omega^{(N_\mathrm{samples})}
    \}$ of the uncertainty realizations and map them to obtain $N_{\mathrm{samples}}$
    market clearing scenarios by evaluating $(P^{\bullet}_{\mathrm{G}}(\omega^{(i)}
    ), P^{\bullet}_{\mathrm{R}}(\omega^{(i)}))$ via the considered polynomial basis.

    Next, to perform the MC calculations, we solve the deterministic Algorithm~\ref{alg:n1_redispatch}
    for all samples, iterating over the market clearing scenarios by setting
    $(p_{\mathrm{G}}^{\bullet}, p_{\mathrm{R}}^{\bullet}) = (P^{\bullet}_{\mathrm{G}}
    (\omega^{(i)}), P^{\bullet}_{\mathrm{R}}(\omega^{(i)}))$
    for each $i \in \mathbb{I}_{N_{\mathrm{samples}}}$ in \eqref{eq:redispatch}.
    By contrast, the PCE-based algorithm requires solving the optimization problem
    \eqref{eq:stoch_redispatch}, which directly considers the forecast distributions.
    Here, the number of times we have to optimize does not depend on $N_{\mathrm{samples}}$,
    since we use the samples only in \eqref{eq:stoch_violation_mc} for the $\varepsilon$-CBCO
    analysis. This results in a significantly smaller computation time ($53.21$ seconds
    vs. $135 5.09$ seconds for the MC approach).

    Table~\ref{tab:stoch_cbco_iterations} provides a detailed look into the iterations
    of the (N-1) stochastic redispatch. Results at iteration 0 correspond to the
    regular (N-0) redispatch, i.e., to the solution of the unmodified problem \eqref{eq:stoch_redispatch}.
    At this stage, the $\varepsilon$-CBCO analysis returns three critical
    outages, the most critical of which is the outage of $(23,25)$, since it
    leads to the overloading of $(26,30)$ resulting in the overall highest mean
    violation calculated via \eqref{eq:mean_violation}. After appending the
    corresponding chance-constraint to \eqref{eq:stoch_redispatch} and re-calculating,
    the situation improves since only one critical outage is detected at
    iteration 1. After modifying \eqref{eq:stoch_redispatch} once again, at
    iteration 2, the solution results in no critical outages. Thus, the execution
    of the algorithm is complete.

    \begin{table}[t]
        \centering

        \caption{CBCO combinations over the iterations of the stochastic (N-1)-secure
        redispatch algorithm.\label{tab:stoch_cbco_iterations}}
        \renewcommand{\arraystretch}{1.2}
        \begin{tabular}{c||m{1cm}|m{1.3cm}|m{1.8cm}|m{1.8cm}}
            \hline
            $i$                & Critical \newline outage & Critical \newline branches & Max. mean \newline violation (MW) & \% of samples \newline with violations \\
            \hline
            \multirow{3}{*}{0} & $(23,25)$                & $\{(26,30)\}$              & 12.8                              & $100$                                  \\
                               & $(26,25)$                & $\{(26,30)\}$              & 11.5                              & $100$                                  \\
                               & $(25,27)$                & $\{(26,30)\}$              & 3.5                               & $99.44$                                \\
            \hline
            1                  & $(26,25)$                & $\{(26,30)\}$              & 1.4                               & $7.23$                                 \\
            \hline
            2                  & $\emptyset$              & $\emptyset$                & 0                                 & 0                                      \\
            \hline
        \end{tabular}
    \end{table}

    \begin{figure*}[t]
        \centering
        \begin{subfigure}
            [b]{0.32\textwidth}
            \centering
            \includegraphics[width=\textwidth]{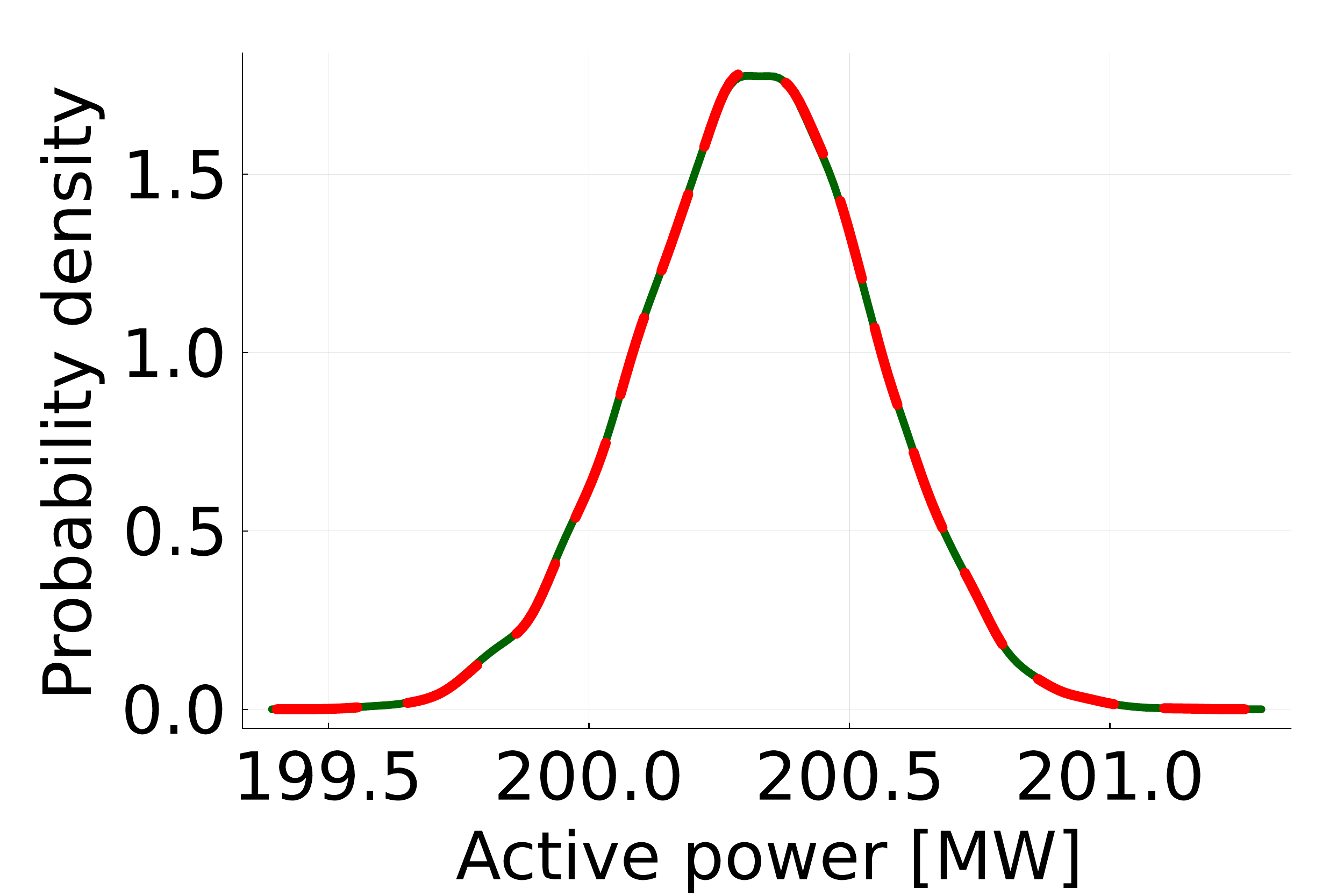}
            \caption{Generator $11 \in \mathcal{G}$}
            \label{fig:A}
        \end{subfigure}\hfill
        \begin{subfigure}
            [b]{0.32\textwidth}
            \centering
            \includegraphics[width=\textwidth]{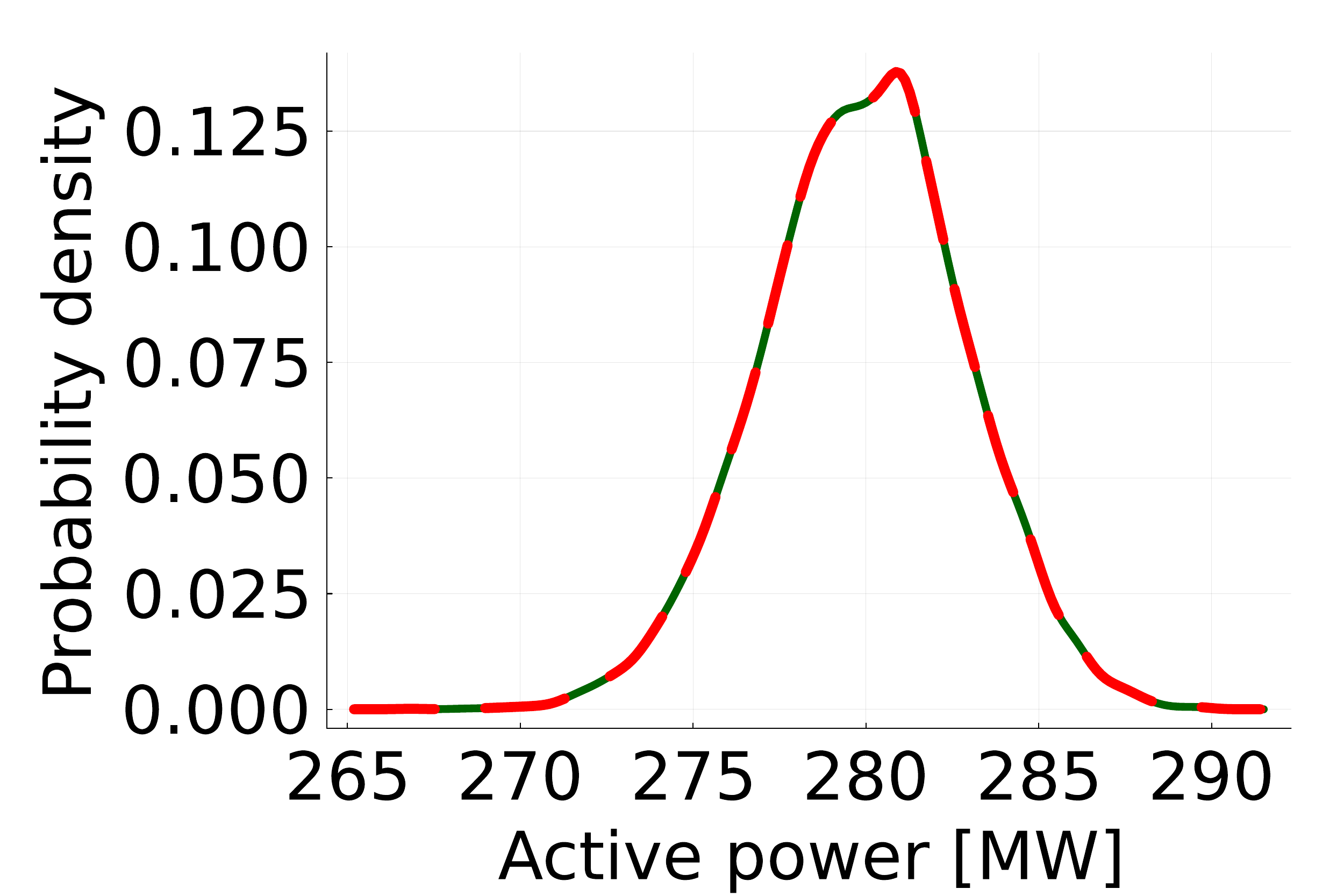}
            \caption{Generator $12 \in \mathcal{G}$}
            \label{fig:B}
        \end{subfigure}\hfill
        \begin{subfigure}
            [b]{0.32\textwidth}
            \centering
            \includegraphics[width=\textwidth]{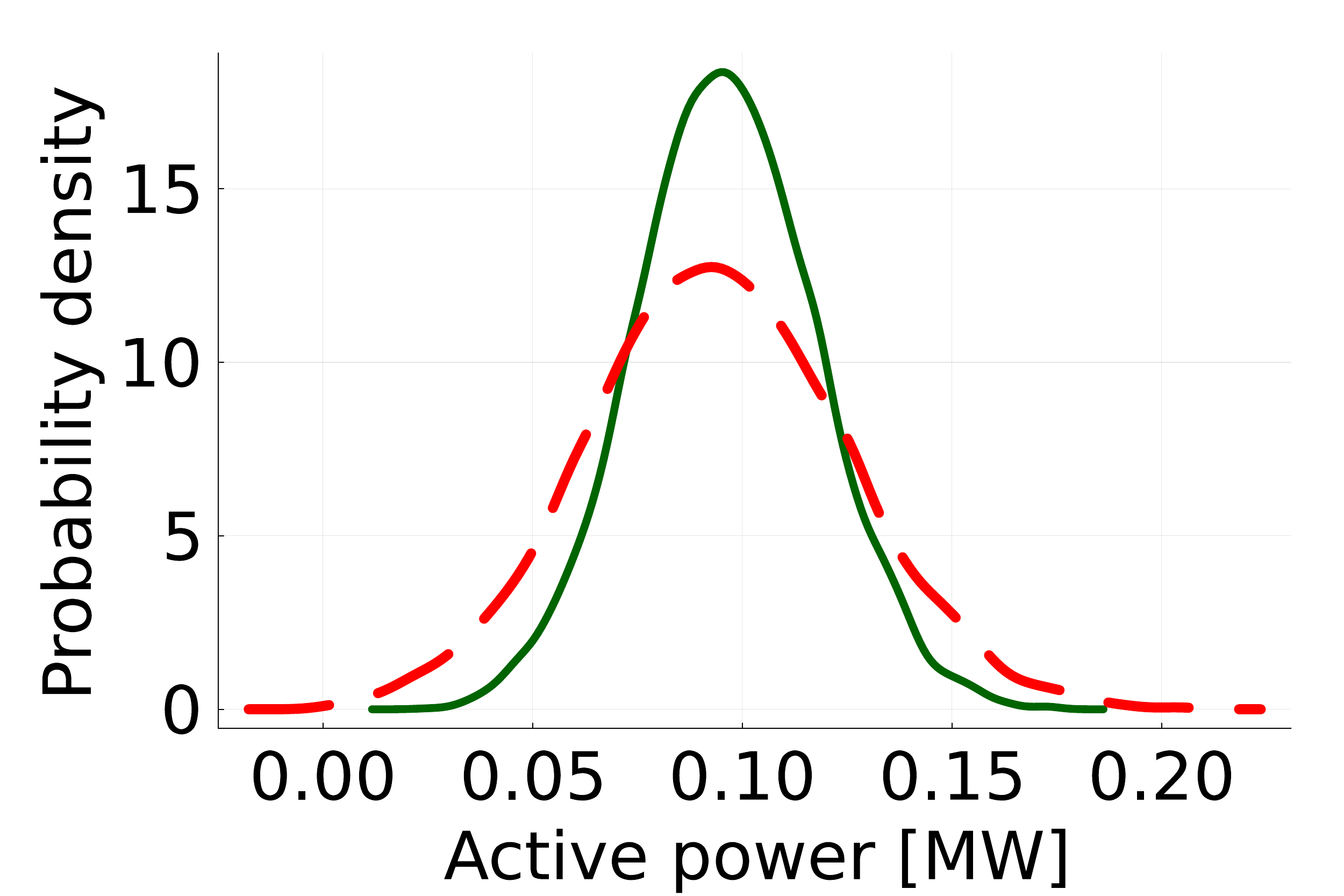}
            \caption{Generator $13 \in \mathcal{G}$}
            \label{fig:C}
        \end{subfigure}
        \caption{Density functions of generator setpoints computed via PCE (\emph{green})
        and MC-based (\emph{red}) (N-1)-secure redispatch.}
        \label{fig:three_pdfs}
    \end{figure*}

    For a more in-depth analysis, we compare the distributions of the solutions
    to the PCE-based and MC-based (N-1)-secure stochastic redispatch. The probability
    density functions obtained by histogram fitting are depicted in Fig.~\ref{fig:three_pdfs}.
    For the sake of brevity, we focus only on a selection of generators
    $\{11,12,13\}\subset \mathcal{G}$ close to the critical branches from Table~\ref{tab:stoch_cbco_iterations}.
    The comparison reveals that for $\{11,12\}\subset \mathcal{G}$ the obtained densities
    are nearly identical, whereas notable differences appear at
    $13 \in \mathcal{G}$ because of its operation at close-to-zero power
    production. Although, the shown densities resemble Gaussian bell curves, due
    to the chosen basis \eqref{eq:uncertainty_source}, they correspond to a linear
    combination of beta-distributed random variables. The similarity of the
    distributions can be certified by comparing the statistical moments of the
    solutions, cf.~Table~\ref{tab:mean_std_comp}.
    \begin{table}[b]
        \centering
        \renewcommand{\arraystretch}{1.2}
        \caption{Comparison between mean and standard derivation (Std) values of
        generator setpoints computed via PCE-based and MC-based (N-1)-secure
        stochastic redispatch.}
        \begin{tabular}{c||l|l|l|l}
            \hline
            $i \in \mathcal{G}$ & \multicolumn{2}{c|}{Mean (MW)} & \multicolumn{2}{c}{Std (MW)} \\
            \cline{2-5}         & PCE                            & MC                          & PCE   & MC    \\
            \hline
            11                  & 200.322                        & 200.323                     & 0.222 & 0.221 \\
            12                  & 279.927                        & 279.927                     & 2.907 & 2.907 \\
            13                  & 0.096                          & 0.094                       & 0.022 & 0.031 \\
            \hline
        \end{tabular}
        \label{tab:mean_std_comp}
    \end{table}

    \section{Conclusions and Outlook}
    \label{sec:conclusion} With the increasing integration of wind and solar
    power---characterized by volatile and non-Gaussian nature---traditional
    redispatch approaches based on deterministic forecasts or MC methods fall
    short in computational efficiency and operational interpretability. In this work,
    we have introduced a PCE-based procedure for stochastic (N-1)-secure
    redispatch. We compare it with the MC approach in an IEEE 118-bus example network.
    The results suggest that our method is able to provide high-accuracy results
    in less computation time. Leveraging the advantages of PCEs, we can provide
    additional interpretability, e.g., by analyzing contributions of the individual
    forecast uncertainties to the operational point of the network based on the PCE
    coefficients of the solution.

    Future work might consider including descrete random-variable decisions to implement
    stochastic unit commitment. In addition, the proposed method could benefit if
    the (usually non-linear) dependencies between individual RES power forecasts
    can be accounted for based on more general descriptions such as, e.g.,
    copulas, rather than correlation matrices.

    \bibliographystyle{ieeetr}
    \bibliography{literature}
\end{document}